\documentclass[fleqn,10pt]{wlscirep}
\usepackage[utf8]{inputenc}
\usepackage[T1]{fontenc}
\usepackage[inkscapelatex=false]{svg}
\usepackage{soul}

\title{The Impact of Large Language Models on K-12 Education in Rural India: A Thematic Analysis of Student Volunteer’s Perspectives}

\author[1]{Harshita Goyal}
\author[1]{Garima Garg}
\author[1]{Prisha Mordia}
\author[2,*]{Veena Ramachandran}
\author[1]{Dhruv Kumar }
\author[1]{Jagat Sesh Challa }

\affil[1]{Dept. of Computer Science \& Information Systems, Birla Institute of Technology \& Science, Pilani, 333031, India}
\affil[2]{Department of Humanities and Social Sciences, Birla Institute of Technology \& Science, Pilani, 333031, India}
\affil[*]{Corresponding author: veena.r@pilani.bits-pilani.ac.in}


\begin{abstract}
AI-driven education, particularly Large Language Models (LLMs), has the potential to address learning disparities in rural K-12 schools. However, research on AI adoption in rural India remains limited, with existing studies focusing primarily on urban settings. This study examines the perceptions of volunteer teachers on AI integration in rural education, identifying key challenges and opportunities. Through semi-structured interviews with 23 volunteer educators in Rajasthan and Delhi, we conducted a thematic analysis to explore infrastructure constraints, teacher preparedness, and digital literacy gaps. Findings indicate that while LLMs could enhance personalized learning and reduce teacher workload, barriers such as poor connectivity, lack of AI training, and parental skepticism hinder adoption. Despite concerns over over-reliance and ethical risks, volunteers emphasize that AI should be seen as a complementary tool rather than a replacement for traditional teaching. Given the potential benefits, LLM-based tutors merit further exploration in rural classrooms, with structured implementation and localized adaptations to ensure accessibility and equity.
\end{abstract}
\begin{document}

\flushbottom
\maketitle

\thispagestyle{empty}

\section{Introduction}\label{sec1}

Rural education in India remains a critical yet persistently under-resourced pillar of the national education system. While India has made considerable strides in expanding access to education over the past two decades, persistent inequities continue to affect rural learners disproportionately. Widespread reforms under landmark initiatives such as the \textit{Sarva Shiksha Abhiyan}\cite{ssa2010}, \textit{Digital India}\cite{digitalindia}, and the \textit{National Education Policy 2020}\cite{nep2020} have aimed to bridge the urban-rural divide, focusing on universal enrollment, digital empowerment, and equitable access to quality learning. Yet, the implementation of these policies often faces bottlenecks at the last mile, where infrastructural fragility, limited digital penetration, and human resource constraints continue to impede meaningful educational transformation.  

Schools in rural areas commonly grapple with inadequate classroom infrastructure, outdated learning materials, teacher shortages, and poor access to basic electricity or internet. According to the Annual Status of Education Report (ASER), a significant proportion of students in rural classrooms do not meet age-appropriate literacy and numeracy benchmarks, often lacking foundational skills in reading and arithmetic \cite{aser2022}. Teachers, when present, are frequently tasked with managing multigrade classrooms, operating without sufficient instructional support or access to evolving pedagogical practices \cite{banerjee2017proof}. These conditions reinforce a cycle of underperformance that contributes to the growing rural-urban education gap, especially in learning outcomes and digital readiness.

In parallel, the rise of Artificial Intelligence (AI)—particularly the advent of Large Language Models (LLMs) like ChatGPT\cite{openai2024}, Bard\cite{bard2024}, and Claude\cite{claude2024}—has transformed global educational paradigms. These models have demonstrated exceptional promise in enabling personalized tutoring, automated assessment, multilingual content delivery, and real-time student feedback at scale \cite{kasneci2023chatgpt, zawacki2019systematic}. In well-resourced urban and private educational environments, LLMs are already being used to scaffold hybrid classrooms, reduce teacher burden, enhance curriculum design, and support individualized learning plans \cite{luckin2016intelligence}. However, the extent to which such technologies can or should be adapted to rural Indian classrooms remains an open and underexplored question. Most prior literature on AI in education has focused either on global north contexts or on elite urban Indian institutions, leaving rural schooling systems conspicuously absent from the empirical record \cite{chaudhury2021personalized}.

This paper addresses that research gap by presenting the grounded perspectives of student volunteers engaged in rural education through national-level service programs like the \href{https://nss.gov.in/}{National Service Scheme (NSS)} and the \href{https://nirmaan.org/}{Nirmaan Organization}. These student volunteers are uniquely situated to provide insight into the potential and pitfalls of deploying AI in resource-constrained rural settings. As urban-educated individuals with direct teaching experience in rural classrooms, they possess a rare dual familiarity with both technological affordances and the day-to-day realities of rural education. Their perspectives offer an invaluable vantage point: they are not only digitally literate and pedagogically aware, but also intimately acquainted with infrastructural limitations, student learning gaps, and the broader sociocultural context that shapes rural schooling.

By foregrounding their lived experiences, this study moves beyond abstract theorization to engage with AI in education as it is perceived "on the ground." We examine how student volunteers assess the role of LLMs—not as a futuristic ideal, but as a tangible possibility for enhancing engagement, reducing instructional load, and bridging content gaps in rural K–12 classrooms. Do they see LLMs as empowering aids, or as technologically incompatible tools in low-connectivity environments? Are they hopeful about the future of AI in education, or cautious due to concerns about bias, over-reliance, and infrastructural feasibility? This study explores these questions through the lens of four research questions:
\\ \\
\noindent \textbf{RQ1:} What are rural student volunteers’ perceptions of AI in K-12 education, and what shapes them?

\noindent \textbf{RQ2:} What are the key barriers to AI adoption in rural schools, including infrastructure, training, and parental attitudes?

\noindent  \textbf{RQ3:} How can AI-driven tools enhance student engagement, personalize learning, and reduce the rural-urban education gap?

\noindent \textbf{RQ4:} What ethical and practical concerns exist around AI in rural education, and how can policies address them?  
\\ \\
To address these questions, we conducted 23 semi-structured interviews with student volunteers teaching in rural schools across Rajasthan and Delhi. These participants offered reflections grounded in their lived experiences, shaped by their dual exposure to urban digital ecosystems and the ground realities of rural education. Their testimonies provide critical insight into how AI can be positioned not as a replacement for teachers or infrastructure, but as a flexible, supportive layer within the existing schooling fabric. Using a qualitative thematic analysis approach based on Braun and Clarke’s (2006) framework \cite{braun2006using}, we identified multiple recurring themes in participant narratives. These include enthusiasm for AI-enabled personalization, frustration with infrastructural bottlenecks, concerns about teacher readiness and resistance, and broader ethical reflections on student over-dependence on automated assistance. Importantly, the perspectives shared by student volunteers reflect a hybrid understanding—technically fluent, but socially rooted—that offers a nuanced roadmap for policy, design, and implementation.

This study contributes novel empirical insights to a discourse that has been dominated by top-down or urban-centric viewpoints. By shifting the focus to the voices of student volunteers who have directly engaged with rural education, we aim to illuminate both the opportunities and constraints of deploying LLMs in marginalized settings. Our findings suggest that while LLMs are not a silver bullet, they hold considerable potential as supplemental educational tools—provided they are contextually adapted, supported by teacher training, and embedded within a framework of equitable digital access and cultural relevance. Through this work, we advocate for a cautious yet proactive exploration of AI integration in rural education ecosystems, grounded in the voices of those working at the frontlines of India's educational challenges.

\section{Related Works}\label{sec2}

\subsection{LLMs and AI in K-12 and Higher Education: A Global and Indian Perspective}
The advent of Large Language Models (LLMs) has significantly influenced global education, with applications spanning from personalized learning assistants to AI-driven curriculum design \cite{brown2020language, bommasani2021opportunities}. Brown et al. \cite{brown2020language} demonstrated GPT-3's capabilities in educational content generation, while Bommasani et al. \cite{bommasani2021opportunities} explored how foundation models can transform educational practices. Research has demonstrated that AI-powered tools can enhance student engagement, automate administrative tasks, and bridge learning gaps \cite{vanlehn2011relative, luckin2016intelligence}. While LLMs have been widely explored in higher education \cite{kasneci2023chatgpt, dillenbourg2016evolution}, their impact on K-12 education, particularly in rural and under-resourced contexts, remains underexplored \cite{chaudhury2021personalized, azubuike2020systematic}. Kasneci et al. \cite{kasneci2023chatgpt} investigated ChatGPT's implications for university teaching and assessment, highlighting both opportunities and challenges.

Research on AI integration in K-12 education has primarily focused on intelligent tutoring systems, adaptive learning environments, and automated grading tools \cite{brusilovsky2007user, woolf2009building}. LLMs have expanded these capabilities by enabling conversational AI tutors, essay evaluation models, and AI-generated instructional content \cite{chen2021multi, winkler2018unleashing}. Chen et al. \cite{chen2021multi} developed multi-modal conversational AI systems that combine text, images, and speech for more engaging K-12 instruction. Prior studies have found that AI-assisted learning platforms can improve knowledge retention, provide real-time feedback, and foster student autonomy \cite{holstein2019co, wang2021student}. However, concerns about AI-generated misinformation, bias in automated assessments, and over-reliance on technology have been raised, particularly in contexts where digital literacy remains low \cite{bender2021dangers, reich2017good}. Bender et al. \cite{bender2021dangers} identified specific risks of LLMs in educational settings, including factual inaccuracies and representational biases.

The success of AI and LLMs in higher education settings has provided insights into their potential scalability in K-12 environments \cite{zawacki2019systematic, hwang2020computational}. Zawacki-Richter et al. \cite{zawacki2019systematic} conducted a systematic review of 146 studies on AI applications in higher education, identifying key success factors for implementation. Universities worldwide have employed AI for automated grading, academic writing support, and plagiarism detection \cite{erickson2020measuring, marciniak2018building}. Research on AI-driven feedback mechanisms suggests that students benefit from personalized AI-generated assessments \cite{wise2018unpacking, kovanovic2020understand}. However, LLM integration in K-12 classrooms requires additional safeguards, particularly in low-resource settings where infrastructure and teacher training gaps persist \cite{rodriguez2021assessing, motz2021pandemic}. Rodriguez-Abitia et al. \cite{rodriguez2021assessing} assessed technological readiness in low-resource schools across multiple countries, highlighting essential prerequisites for AI adoption.

In the Indian context, there has been a growing focus on AI-driven education, with government initiatives like Diksha and Swayam emphasizing technology-enhanced learning \cite{chauhan2017overview, agarwal2020technology}. Studies have explored AI's potential in personalized learning for linguistically diverse classrooms \cite{chandwani2021extending} and AI-assisted educational content generation for regional languages \cite{raghavan2020adapting}. Chandwani et al. \cite{chandwani2021extending} developed and tested adaptive learning systems for multilingual classrooms in six Indian states, documenting significant improvements in learning outcomes. However, researchers have identified significant challenges in AI adoption, including low teacher familiarity with AI tools, infrastructural limitations, and digital divide concerns \cite{narayan2019challenges, siddiqui2020closing}. The reluctance of educators to embrace AI without adequate training has been widely reported \cite{muralidharan2019disrupting}, necessitating structured AI literacy programs. Recent studies have addressed AI implementation in rural Indian education. Tripathi et al.~\cite{tripathi2025leveraging} highlighted AI's potential to bridge the urban-rural educational divide by providing otherwise unavailable learning opportunities. Similarly, Sharma et al.~\cite{sharma2024ai} examined AI as a catalyst for educational equity, identifying rural-urban disparities and proposing targeted AI interventions to address these inequalities.\cite{goel2024ai} did literature review of the differet ways AI can be used in education in India. 

\subsection{Sociological Perspectives and Challenges of AI Adoption in Rural Indian Education}
From a sociological standpoint, educational disparities in rural India have been extensively studied, with scholars emphasizing economic, linguistic, and infrastructure-related challenges \cite{desai2010human, dreze2001school}. Desai et al. \cite{desai2010human} conducted a comprehensive human development survey across 41,554 households in India, documenting stark rural-urban educational divides across multiple indicators. Research has documented the rural-urban learning divide, teacher shortages, and the role of socioeconomic status in academic achievement \cite{banerjee2017proof, kumar2018how}. Studies on digital interventions in rural education have demonstrated mixed success, highlighting that technology alone cannot overcome pedagogical and infrastructural gaps \cite{pal2016my, chatterjee2020comparative}. Pal et al. \cite{pal2016my} followed 28 rural schools implementing technology-based learning over three years, finding that outcomes depended heavily on teacher support and infrastructure quality. AI-driven education must, therefore, be implemented with localized pedagogical adaptations, community engagement, and sustained teacher support \cite{arora2019next, mitra2017equal}.

Despite the promise of AI-powered education, multiple barriers hinder large-scale adoption in rural India. Research indicates that infrastructure deficiencies, digital illiteracy among teachers, and financial constraints remain primary obstacles \cite{gulati2008technology, pal2018fallacy}. Additionally, studies on AI in multilingual education suggest that language models predominantly trained in English fail to cater to regional dialects and cultural contexts \cite{das2021developing, choudhury2019ai}. Das et al. \cite{das2021developing} developed and evaluated NLP models for low-resource Indian languages, documenting performance gaps compared to English models. Ethical concerns, including algorithmic bias, data privacy, and the risk of over-reliance on AI-generated content, have also been widely debated in the literature \cite{mohamed2020decolonial, warschauer2010new}.

AI and LLMs hold the potential to reduce educational disparities by providing quality digital learning resources to underprivileged students \cite{reich2020failure, nye2015intelligent}. Reich \cite{reich2020failure} analyzed failed ed-tech interventions to extract principles for more equitable implementation, particularly in resource-constrained settings. Studies have shown that AI-driven personalized learning environments can enhance academic performance in rural settings, provided adequate teacher training and infrastructural support are available \cite{kizilcec2020scaling, naik2019learning}. However, critics caution that the digital divide may be exacerbated if AI-based learning remains concentrated in well-funded institutions \cite{selwyn2020human, rafalow2020digital}. Research advocates for hybrid learning models where AI supplements, rather than replaces, traditional teacher-led instruction \cite{dillenbourg2022classroom, luckin2016intelligence}. Dillenbourg et al. \cite{dillenbourg2022classroom} developed frameworks for teacher-AI collaboration that maintain human pedagogical leadership while leveraging AI capabilities.

\subsection{Ethical Considerations in AI-Driven Education}
Studies on AI ethics in education have emphasized the need for transparent data governance, accountability in AI-generated content, and policies that prevent algorithmic biases \cite{prinsloo2018student, zawacki2019systematic}. Researchers have highlighted concerns regarding student privacy, AI misinformation, and the potential for AI-generated content to limit critical thinking skills \cite{knox2020artificial, holmes2022ethics}. Knox et al. \cite{knox2020artificial} critiqued the anthropomorphization of AI in educational contexts, warning about its impact on student agency and critical engagement. Furthermore, evidence suggests that students who rely excessively on AI for academic tasks may develop weaker problem-solving skills \cite{sweller2019cognitive, kim2021impact}. Sweller et al. \cite{sweller2019cognitive} extended cognitive load theory to AI-assisted learning, identifying scenarios where automation might impede rather than enhance learning. Ethical AI adoption frameworks stress the importance of teacher involvement, policy oversight, and AI adaptations tailored to specific learning environments \cite{selwyn2019whats, zimmerman2020designing}.

The role of LLMs in K-12 education in rural India represents an emerging yet complex research area. While AI-powered tools offer opportunities for enhancing education equity, significant infrastructural, pedagogical, and ethical challenges remain. Prior research underscores the need for localized AI training programs, teacher-AI collaboration models, and policy-driven AI integration strategies \cite{pedro2019artificial, popenici2017exploring}. Future work should focus on longitudinal studies evaluating AI's real-world impact on rural education and developing AI solutions that align with India's multilingual, socioeconomically diverse education system \cite{shah2019digital, pal2017accessibility}. These studies will need to address the unique challenges of implementing AI in contexts where traditional educational infrastructure may be limited, and where cultural and linguistic diversity requires thoughtful adaptation of AI tools designed primarily for Western educational contexts.

\subsection{Positioning Our Work in the Existing Literature}
While prior research has explored the role of AI and LLMs in education, most studies focus on urban contexts, higher education, or well-resourced environments. There remains limited empirical work that captures the ground-level realities of rural schools in India, particularly from the lens of volunteer teachers who navigate both urban AI exposure and rural infrastructural constraints. The existing literature frequently examines AI implementation in education from a top-down perspective, focusing on technological capabilities and theoretical benefits rather than practical implementation challenges in resource-constrained settings. Additionally, many studies evaluate AI educational tools in controlled environments that do not reflect the complex socio-economic realities of rural Indian education. This disconnect between theoretical potential and practical application represents a critical gap in the current research landscape.

Our study addresses this gap by focusing on the perceptions of student volunteers working directly with underprivileged K-12 learners in rural settings. Unlike most evaluations of AI tools, which center on system performance or classroom deployment in controlled environments, we foreground the socio-cultural, infrastructural, and pedagogical barriers as perceived by volunteer educators embedded in rural outreach. This ground-up approach provides valuable insights into the practical challenges of AI implementation that may not be apparent in more controlled research settings. Moreover, we propose a context-aware, AI-driven solution—tailored for NCERT\cite{ncert2025}-based rural curricula—that emerges directly from the thematic concerns raised by the volunteers.

By developing solutions based on identified needs rather than predetermined technological capabilities, our approach aligns with recommended practices for sustainable educational technology implementation. In doing so, our work not only adds to the literature on AI in education but also grounds it in the lived realities of India's rural schooling ecosystem, potentially offering a more viable pathway for meaningful AI integration in similar contexts worldwide.

\begin{table}[ht]
\centering
\caption{Comparison of Existing Works and Our Contribution}
\label{tab:litcomparison}
\begin{tabular}{|p{4.5cm}|p{5.5cm}|p{5.5cm}|}
\hline
\textbf{Aspect} & \textbf{Existing Literature} & \textbf{Our Contribution} \\ \hline

\textbf{Target Educational Level} & Primarily higher education and urban K–12 \cite{kasneci2023chatgpt, zawacki2019systematic} & Focus on rural K–12 settings across Rajasthan and Delhi \\ \hline

\textbf{Technology Focus} & AI, ITS, LLMs, adaptive learning platforms \cite{vanlehn2011relative, woolf2009building, chen2021multi} & Large Language Models (LLMs) like ChatGPT, Bard, and Claude, assessed from a grassroots lens \\ \hline

\textbf{Contextual Setting} & Urban or developed regions, controlled deployments \cite{luckin2016intelligence, chandwani2021extending}with emerging research on rural contexts\cite{tripathi2025leveraging,sharma2024ai} & Real-world rural education settings with infrastructural constraints \\ \hline

\textbf{Methodological Approach} & Top-down evaluation of AI tools; system performance metrics \cite{brusilovsky2007user, wang2021student} & Grounded qualitative study via 23 semi-structured interviews with student volunteers \\ \hline

\textbf{Key Stakeholders} & Teachers, students in well-resourced schools/universities \cite{dillenbourg2016evolution, erickson2020measuring} & Student volunteers with dual exposure to urban AI tools and rural pedagogy \\ \hline

\textbf{Identified Barriers} & Bias, misinformation, over-reliance on AI \cite{bender2021dangers, sweller2019cognitive} & Adds infrastructural bottlenecks, digital divide, and cultural mismatch in rural India \\ \hline

\textbf{Policy Integration} & Mostly theoretical or pilot-based recommendations \cite{rodriguez2021assessing} & Context-aware AI integration strategies based on lived volunteer experiences \\ \hline

\textbf{Ethical Framing} & Privacy, cognitive overload, algorithmic bias \cite{knox2020artificial, selwyn2019whats} & Emphasizes teacher readiness, AI misuse risks, and localized AI literacy needs \\ \hline

\end{tabular}
\end{table}

\section{Methodology}

To explore how Large Language Models (LLMs) may impact K-12 education in rural India, we conducted a qualitative study using semi-structured interviews with 23 student volunteer teachers. These participants were engaged through educational outreach programs in Rajasthan and Delhi, and their insights were thematically analyzed following Braun and Clarke’s framework. The methodology and recruitment process are detailed in the subsections below.

\subsection{Study Design}

\subsubsection{Procedure}

Our study procedure involved two phases: First, participants were provided with an explanation of our project and the scope of generative AI in rural education. Second, semi-structured interviews were conducted to explore their perceptions and experiences regarding AI’s role in rural education. The study aimed to investigate how volunteer teachers perceive the integration of generative AI in rural education, particularly in addressing learning gaps, supporting teachers, and overcoming infrastructure constraints. The research design was developed to ensure both depth and flexibility, allowing participants to provide detailed responses while also maintaining consistency across interviews.

\subsubsection{Semi-Structured Interviews}

A total of 23 interviews were conducted, each lasting 20–25 minutes. Interviews were conducted online via video or audio calls to ensure accessibility and convenience for participants. The semi-structured format enabled a balance between guided inquiry and open-ended discussion, ensuring that core themes were covered while also allowing participants to elaborate on issues they found significant.

Each interview followed a structured process:
\begin{itemize}
    \item \textbf{Consent \& Introduction}: The interviewer provided an overview of the study and obtained verbal consent for participation and recording.
    \item \textbf{General AI Awareness \& Exposure}: Participants were asked about their familiarity with generative AI tools and their experiences using digital learning platforms.
    \item \textbf{Perceptions of AI in Rural Education}:
    \begin{itemize}
        \item Their perspectives on AI’s role in supporting student learning.
        \item Potential benefits and drawbacks of AI-driven tutoring systems.
        \item Practical challenges to AI adoption in rural education, such as infrastructure and teacher training.
        \item Ethical considerations, including concerns about AI accuracy and digital accessibility.
    \end{itemize}
    \item \textbf{Emergent Probing}: Additional questions were introduced based on participant responses to capture nuanced insights.
\end{itemize}

Interviews were conducted in English, recorded with participant consent, and later transcribed for analysis. Given the qualitative nature of the study, the interviews served as a primary source of data to understand the attitudes, expectations, and concerns of volunteer teachers regarding AI in rural education. Additionally, efforts were made to ensure participant comfort and encourage candid responses. Interviewers maintained a conversational tone and assured participants that there were no right or wrong answers. This helped mitigate response biases and ensured that participants openly shared their authentic views. The collected data was subsequently analyzed using thematic analysis, following Braun \& Clarke’s (2006) framework, to identify recurring patterns and themes across participant responses.

\subsection{Participants and Recruitment}

\subsubsection{Participant Demographics}

A total of 23 volunteer teachers participated in this study. Unlike full-time educators, these volunteers teach as part of their university-run social service programs and do not have formal pedagogical training. However, their experiences bridging urban and rural education systems make them valuable informants. Most of these volunteers primarily teach in rural schools in Pilani, Rajasthan, while 1–2 participants teach in schools in Delhi.

\begin{table}[h]
    \centering
    \begin{tabular}{ll}
        \toprule
        \textbf{Category} & \textbf{Details} \\
        \midrule
        \textbf{Gender} & Female: 8, Male: 15 \\
        \textbf{Age (years)} & Min: 20, Max: 25, Avg: 22.5 \\
        \textbf{Teaching Experience (years)} & Min: 2, Max: 4, Avg: 3.0 \\
        \textbf{Familiarity with AI Tools} & High: 13, Moderate: 10, Low: 0 \\
        \textbf{Technology Use in Teaching} & Mobile phones: 19, Laptops: 8, Offline teaching only: 3 \\
        \bottomrule
    \end{tabular}
    \caption{Demographic and Technological Background of Participants}
    \label{tab:participant_info}
\end{table}

\subsubsection{Justification for Selecting Volunteer Teachers}

Volunteer teachers were selected over full-time educators due to their unique positioning between urban and rural education systems. Many of these volunteers had exposure to AI-driven educational tools in their personal or academic settings, allowing them to critically assess the feasibility of such tools in rural schools. Additionally, as these volunteers frequently work with underprivileged students, they provide valuable insights into how AI can support learning in low-resource environments.

\subsubsection{Recruitment Challenges}
While NSS and Nirmaan facilitated participant recruitment, several challenges emerged during the process. Many volunteers possessed minimal prior exposure to AI tools, necessitating additional contextual briefings before interviews could proceed effectively. Scheduling interviews proved particularly difficult as participants juggled competing demands from their academic studies and other volunteer commitments. Furthermore, we encountered initial skepticism among some participants regarding AI's potential impact on education, requiring careful explanation of our study objectives to establish trust and encourage open dialogue. These recruitment challenges ultimately informed our methodological approach, as we allocated additional time for preliminary discussions and maintained flexible scheduling to accommodate participants' availability.

\subsection{Data Collection and Analysis}

The collected data was analyzed using thematic analysis, following Braun \& Clarke’s (2006) \cite{braun2006using} framework. The analysis process began with transcription and familiarization, where audio recordings were transcribed verbatim, and researchers reviewed the transcripts multiple times to gain a thorough understanding of the data and identify initial patterns. This was followed by initial coding, where open coding was applied to label significant phrases and statements, assigning codes to key concepts such as "lack of digital access," "teacher skepticism toward AI," and "AI as an adaptive learning tool." Two researchers independently coded the data, and inter-coder reliability was established through consensus discussions. 

\subsection{Ethical Considerations}

Throughout the study, we carefully addressed ethical concerns for transparency and participant privacy. Our materials and
procedures were approved by the university’s Institutional Human Ethics Committee (IHEC), BITS Pilani, Rajasthan and were carried out in accordance with relevant
guidelines and regulations. In surveys, we highlighted voluntary participation, response confidentiality, and the study’s purpose.
The survey introduction also explained the participants’ role in contributing to the study and assured them of the anonymity of
their responses, as no names or identifying information was collected. Before interviews, participants provided explicit written
and verbal informed consent, assuring anonymity. Thus informed consent was obtained from all subjects. Data was anonymized,
and securely stored with limited team access. 

\subsection{Limitations}

While this study provides valuable insights, several limitations must be acknowledged. One primary limitation is participant bias, as all respondents were volunteer teachers rather than full-time educators. Their perspectives may differ from those of government school teachers who have long-term exposure to rural education and institutional policies. Additionally, all interviews were conducted remotely due to logistical constraints. An ethnographic approach or field study could provide deeper insights into how AI is practically integrated into rural classrooms. Despite these limitations, the study offers a foundational understanding of volunteer teachers’ perceptions of AI in rural education and highlights areas for further exploration.

This methodology outlines a rigorous approach to exploring the role of AI in rural education from the perspective of volunteer teachers. By conducting semi-structured interviews and using thematic analysis, the study captures both the potential and challenges of AI-based learning in low-resource environments. The findings from this study offer practical insights for policymakers, NGOs, and AI developers, highlighting the need for targeted interventions to make AI-driven education accessible and effective in rural schools. Future research should focus on pilot implementations of AI tools and their impact on student learning outcomes.

\begin{figure}[h!]
    \centering
    \includegraphics[width=1\textwidth]{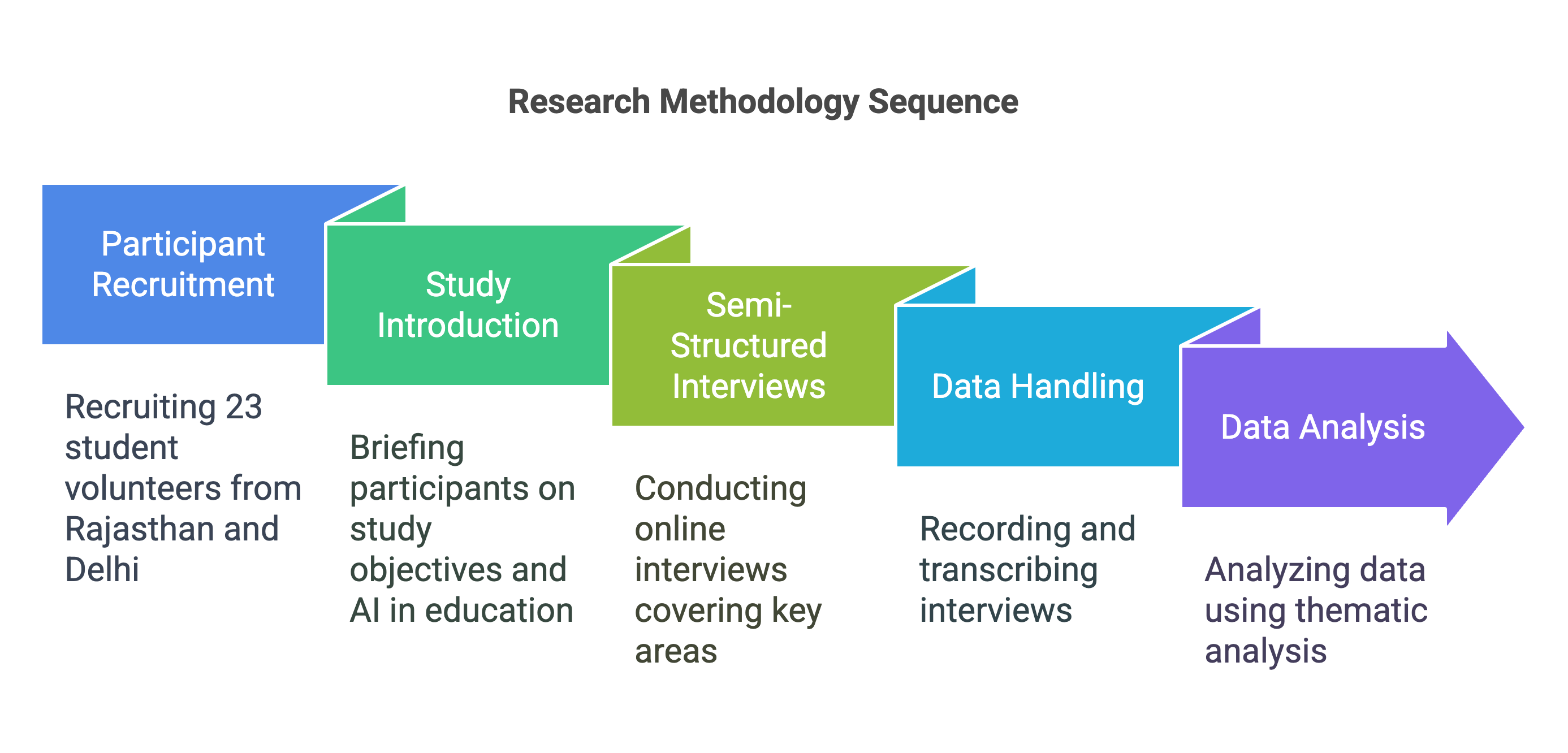}
    \caption{Research Methodology Flowchart}
    \label{fig:research-methodology-flow}
\end{figure}

The overall research methodology is summarized in the diagram provided in Figure~\ref{fig:research-methodology-flow}.

\section{Findings}
Our findings indicate that rural educators have varying levels of awareness and familiarity with AI tools, with many expressing both optimism and skepticism regarding their integration into classrooms. While participants recognized AI's potential to personalize learning, reduce workload, and enhance student engagement, others were concerned about over-reliance on AI, lack of infrastructure, and the digital divide. 

The study revealed that student volunteers see AI as a supplementary tool rather than a replacement for traditional teaching, emphasizing the need for human oversight, proper training, and context-specific AI implementation. Many noted the importance of blended learning models, where AI supports but does not overshadow critical thinking and teacher-student interactions. Furthermore, key barriers such as cost, language accessibility, and parental skepticism must be addressed for effective AI adoption in rural K-12 education.

In the following sections, we explore these themes in depth, beginning with our volunteers' perceptions of AI (Section~\ref{sec:challenges}), followed by the potential benefits of LLMs (Section~\ref{sec:role}), the challenges and ethical considerations (Sections~\ref{sec:personalized} and \ref{sec:infra}), and concluding with infrastructure, policy recommendations, and future directions (Sections~\ref{sec:ethics} and \ref{sec:community}).

\subsection{Challenges in Rural Education}
\label{sec:challenges}

The integration of AI and Large Language Models (LLMs) in rural K-12 education faces several challenges, primarily rooted in infrastructure deficits, teacher quality, student engagement, and parental attitudes. While AI has the potential to bridge educational gaps, its adoption remains restricted by these structural barriers.

A major obstacle is the lack of infrastructure, including unreliable internet, limited access to digital devices, and underfunded schools. Among the 23 participants interviewed, 15 emphasized that basic facilities were inadequate, making AI-based learning tools inaccessible. The Digital Divide Theory \cite{vandijk2020digital} explains that disparities in access reinforce existing socio-economic disadvantages. Without foundational investment, AI will likely remain an urban phenomenon. 
\begin{center}\textit{"Infrastructure is a huge issue—basic necessities like classrooms, washrooms, and even clean water are sometimes lacking."} [P12]\end{center}

When asked about their familiarity with AI and its integration into teaching, participants highlighted lack of technical knowledge amongst teachers as a major barrier. Many rural teachers are unfamiliar with AI and related digital tools, making effective integration challenging. 
\begin{center}\textit{"I don't think most government school teachers are aware of GenAI yet."} [P4]\end{center}
Additionally, a shortage of qualified teachers exacerbates the issue, with one participant noting that a government school with a computer lab had no computer teacher for over two years. Out of the 23 interviewees, 13 expressed concerns over their limited access to AI training programs. Without structured training, even teachers who recognize AI's potential struggle to implement it in classrooms.

Student engagement remains a persistent challenge, with low attendance, lack of motivation, and foundational learning gaps affecting AI implementation. Ryan and Deci's Self-Determination Theory \cite{ryan2000self} suggests that students require autonomy, competence, and relatedness to remain engaged. However, 12 participants reported that students in rural areas struggle with literacy and numeracy, making AI-driven learning tools less effective. 
\begin{center}\textit{"Often, the students in lower grades aren't performing at grade level, with 1st graders, for example, sometimes being at the level of preschool."} [P21]\end{center}
AI could provide adaptive learning pathways, but its impact is limited without strong foundational education.

Parental skepticism toward AI and digital education presents another significant barrier. Many rural families prioritize traditional learning methods and may perceive AI as unnecessary or distracting. Venkatesh et al.'s Unified Theory of Acceptance and Use of Technology (UTAUT) \cite{venkatesh2003user} highlights parental support as crucial for student engagement with technology. Among the respondents, 9 noted that convincing parents about AI's benefits remains an uphill task. 
\begin{center}\textit{"Some parents might oppose AI tools, but it is our role to explain how it enhances their child's learning experience."} [P15]\end{center}
Without parental buy-in, AI-driven education in rural areas will struggle to gain traction.

These findings indicate that while AI offers opportunities for improving rural education, its successful implementation depends on addressing core deficits in infrastructure, teacher training, student preparedness, and parental engagement. Without systematic interventions, AI adoption risks deepening the educational divide rather than bridging it.

\subsection{The Role of Technology in Education}
\label{sec:role}

Our findings indicate that while rural K-12 schools in India face significant limitations in technology adoption, there is a growing awareness among educators about the potential of AI and LLMs to enhance learning experiences. However, integration remains inconsistent due to digital access disparities, lack of teacher training, and skepticism toward AI's role in education.

When asked about the availability of digital tools, participants widely acknowledged the limited access to technology in rural schools. While some schools in Pilani have computer labs that introduce students to basic web development, most government schools restrict technology use to administrative tasks rather than classroom instruction. Among the 23 participants, 15 noted that internet access in their schools was unreliable, making it difficult to integrate AI-based tools. 
\begin{center}\textit{"Private schools have introduced some Gen AI and tech-based tools, but government schools mostly stick to traditional, book-based methods."} [P19]\end{center}
\begin{center}\textit{"Most schools don't have reliable internet, and tech access is limited."} [P10]\end{center}
Even though mobile phones are prevalent in households, their use for educational purposes remains minimal, with students mainly using them for entertainment or informal learning through YouTube.

AI is increasingly seen as a valuable learning aid, particularly in addressing educational gaps between urban and rural students. Participants highlighted how AI-powered tools could offer personalized learning experiences, real-time feedback, and adaptive lesson plans. 
\begin{center}\textit{"Generative AI could be incredibly useful. Tools like ChatGPT could make it easier for students to access knowledge without needing to go to a teacher every time they have a question."} [P1]\end{center}
However, concerns about students developing over-reliance on AI instead of honing problem-solving skills were also raised. Despite these concerns, AI's ability to supplement learning through accessible, interactive content remains promising. The Zone of Proximal Development \cite{vygotsky1978mind} suggests that students benefit most when they receive scaffolded support, indicating AI could be effective in reinforcing learning when guided by teachers.

Volunteers emphasized that technology should be integrated into a hybrid approach rather than replace traditional instruction. While AI can assist in knowledge dissemination, human teachers remain essential for imparting critical thinking skills and emotional intelligence. 
\begin{center}\textit{"A balanced approach—not completely technology-driven and not entirely human-based—would be very supportive for kids in rural areas."} [P14]\end{center}
\begin{center}\textit{"NGOs and volunteers can help train teachers in rural schools to integrate AI tools effectively."} [P18]\end{center}
Among the respondents, 10 felt that without proper teacher training, AI's potential in education would remain underutilized.

LLMs provide new opportunities for enhancing conceptual learning through interactive content. AI-generated visualizations, real-world simulations, and explanatory videos can simplify complex subjects, particularly in STEM education. 
\begin{center}\textit{"AI-generated videos help students understand better because visuals stick in their memory."} [P12]\end{center}
\begin{center}\textit{"Teachers could use GenAI tools like ChatGPT to generate images and videos for better concept visualization."} [P18]\end{center}
Cognitive Load Theory \cite{sweller1988cognitive} suggests that students learn best when complex information is presented in multimodal formats, reinforcing the potential of AI-driven visualization tools to enhance student engagement and comprehension.

While AI holds promise for transforming rural education, fundamental challenges such as digital infrastructure, teacher preparedness, and accessibility must be addressed. A well-planned integration of AI tools, combined with human oversight, will be essential in ensuring equitable learning opportunities for students in rural areas.

\subsection{Personalized Learning with GenAI}
\label{sec:personalized}

Building on the discussion of AI's role in education, this section explores how GenAI can support personalized learning and address individual student needs. AI-driven tools can cater to individual student needs, support teachers, bridge knowledge gaps, and help reduce the rural-urban education divide.

When asked about the potential of AI in individual learning, participants highlighted its ability to cater to students at different levels of comprehension. Among the 23 participants, 14 noted that AI could help address foundational learning gaps by tailoring content to students' needs. Some have experimented with adaptive learning tools where difficulty levels adjust based on student responses. 
\begin{center}\textit{"I've tried out some tools for customized learning, where the difficulty adjusts based on responses."} [P17]\end{center}
Others emphasized how AI could make learning more engaging by generating personalized content. 
\begin{center}\textit{"Gen AI could assist by creating stories, rhymes, or other engaging formats that make learning enjoyable, especially for younger kids."} [P15]\end{center}
Additionally, AI could provide foundational support for struggling students, reinforcing basic skills such as reading, writing, and sentence formation. Vygotsky's Zone of Proximal Development \cite{vygotsky1978mind} suggests that students benefit most when instructional support is scaffolded, indicating that AI-driven tools could be particularly effective in bridging these gaps.

Volunteers also recognized that AI could help reduce the teacher's workload while improving the overall teaching experience. Many noted that AI can act as a support tool rather than a replacement, making instruction more efficient. 
\begin{center}\textit{"AI would take a lot of the burden off teachers and make their workload easier."} [P16]\end{center}
AI-powered chat systems can also provide real-time guidance for educators, helping them refine their teaching strategies. 
\begin{center}\textit{"Chat-based systems can help teachers understand how to explain concepts better to students."} [P20]\end{center}
To ensure effective AI adoption, several respondents suggested conducting workshops and training programs in rural areas. 
\begin{center}\textit{"NGOs and volunteers can help teachers by training them in rural schools to familiarize them with AI tools."} [P3]\end{center}
Among the respondents, 11 highlighted that teacher training is crucial for the successful integration of AI.

GenAI has the potential to bridge knowledge gaps by providing students with immediate access to educational support. AI tools allow students to learn at their own pace, request simplified explanations, and receive answers in a structured manner. 
\begin{center}\textit{"Gen AI tools are great learning tools, especially if you're new to a topic. You can ask questions, request simpler explanations, and go at your own pace."} [P21]\end{center}
Another emphasized the importance of local language support:
\begin{center}\textit{"AI's multilingual capabilities could help teachers and students communicate better, acting as a translator where needed."} [P6]\end{center}
AI's ability to break down complex topics into digestible lessons makes it an effective tool for reducing the educational divide. However, one participant also noted a possible drawback:
\begin{center}\textit{"At the same time, I feel AI sometimes kills my creativity, as I depend on it to find answers instead of thinking on my own."} [P15]\end{center}
This highlights the risk of over-reliance on AI, where students might prioritize convenience over developing independent problem-solving skills if not guided properly. Cognitive Load Theory \cite{sweller1988cognitive} suggests that students learn better when information is presented in multiple formats, reinforcing the role of AI in improving comprehension through structured, simplified explanations.

Reducing the urban-rural education gap remains a key challenge, but AI integration can play a transformative role in leveling the playing field. Several participants pointed out that round-the-clock access to AI-powered education could inspire rural students to take education more seriously. 
\begin{center}\textit{"These tools would make knowledge accessible 24/7. If rural students could access the same resources as urban ones, they might become more passionate about their education and career aspirations."} [P10]\end{center}
AI-powered platforms could also introduce students to career opportunities beyond traditional rural roles. 
\begin{center}\textit{"Successful integration of Gen AI in rural schools would mean students gaining comfort with technology, recognizing its potential, and exploring careers beyond traditional roles."} [P9]\end{center}
Among the participants, 12 emphasized that AI-driven tools could provide rural students with better insights into career paths and technological advancements.

While GenAI presents significant opportunities for rural education, ensuring proper implementation is crucial. Participants emphasized that AI should be used as a learning aid rather than a complete replacement for human educators. With adequate infrastructure, structured teacher training, and well-planned adoption strategies, GenAI could revolutionize personalized learning in rural India, fostering a more inclusive and effective education system.

\subsection{Barriers to GenAI Adoption}
\label{sec:infra}

While GenAI holds promise for transforming rural education, several barriers hinder its widespread adoption. Cost and accessibility, language barriers, infrastructure limitations, and inadequate teacher training emerged as key challenges.

When asked about the feasibility of integrating AI tools, participants frequently cited cost and accessibility as a primary concern. Among the 23 participants, 11 noted that affordability of devices and data plans remains a challenge. While mobile internet access is improving, many families cannot afford the necessary hardware or consistent internet access. 
\begin{center}\textit{"Many families can't afford consistent data or even a device for every child."} [P4]\end{center}
\begin{center}\textit{"The first barrier might be the cost of application, as rural schools won't have enough computer or internet facilities."} [P18]\end{center}
Without financial investment in infrastructure, AI-driven learning will remain out of reach for many rural students, reinforcing the digital divide. The Capability Approach \cite{sen1999development} highlights that access to technology alone is not enough; equitable learning opportunities require financial, infrastructural, and social support to ensure meaningful engagement.

Language presents another significant challenge in GenAI adoption. Many AI tools primarily operate in English or Hindi, whereas rural students are more comfortable in regional languages. 
\begin{center}\textit{"Language is a big barrier; students struggle with English, especially since the terms in textbooks, like those in NCERT, are in English."} [P20]\end{center}
\begin{center}\textit{"AI tools must be available in local languages, as students often struggle to understand English."} [P19]\end{center}
While some startups are working on AI solutions in regional languages, widespread accessibility remains limited. If AI tools fail to accommodate local linguistic needs, their educational impact will be significantly diminished.

Infrastructure challenges further complicate AI adoption. Many rural schools lack the necessary technological resources, such as reliable internet and functional computer labs. Among the respondents, 8 highlighted that connectivity issues frequently disrupt digital learning. 
\begin{center}\textit{"The biggest challenge is that rural schools lack proper infrastructure—good computers and stable internet are missing."} [P11]\end{center}
\begin{center}\textit{"Lack of internet connectivity is a huge problem for implementing GenAI in rural education."} [P8]\end{center}
Even when digital tools are introduced, unstable internet access and outdated equipment limit their effectiveness. Improving infrastructure, including stable internet, well-equipped computer labs, and AI-compatible devices, is essential for successful GenAI integration in rural education.

A significant barrier to GenAI adoption is the lack of teacher training. Many participants in rural areas are unfamiliar with AI tools and resistant to their use. 
\begin{center}\textit{"Teachers often resist AI tools like ChatGPT, seeing them as a shortcut or a form of cheating rather than an educational aid."} [P23]\end{center}
\begin{center}\textit{"Teaching teachers to use AI tools in rural areas will be a tough task."} [P2]\end{center}
Training initiatives led by NGOs and volunteers could help bridge this gap by equipping teachers with the necessary digital literacy skills. Without structured teacher training, AI implementation will remain fragmented and ineffective.

These barriers highlight the complexities of integrating GenAI into rural education. While AI offers significant opportunities, overcoming financial, linguistic, infrastructural, and pedagogical challenges is crucial to ensuring equitable access and long-term adoption. Participants also cautioned that without careful implementation, AI could widen the existing educational divide rather than bridging it.

\subsection{Ethical and Practical Concerns}
\label{sec:ethics}

While GenAI presents new opportunities for rural education, it also raises ethical and practical concerns. Participants highlighted risks such as overdependence on AI, data privacy issues, potential misuse, and cultural resistance to adopting AI-driven learning tools.

When asked about AI's role in student learning, participants expressed concerns about overdependence. Some worried that students might prioritize convenience over deep learning, relying on AI for quick answers rather than engaging critically with content. Among the 23 participants, 8 mentioned concerns related to AI reducing students' creativity and critical thinking. 
\begin{center}\textit{"The creativity of students could be disrupted. Before GenAI, students would delve deep into discussions and write essays. With AI, they might just type and get answers without much thinking."} [P16]\end{center}
\begin{center}\textit{"At the same time, I feel AI sometimes kills my creativity, as I depend on it to find answers instead of thinking on my own."} [P5]\end{center}
Self-Determination Theory \cite{ryan2000self} suggests that for students to develop autonomy in learning, they need to engage with material actively rather than passively consuming information. Ensuring structured learning environments where AI supports, rather than replaces, critical thinking is crucial.

Data privacy concerns were also raised, particularly regarding the security of student information and uploaded content. Some participants expressed hesitation about the nature of data being shared with AI tools, emphasizing the need for proper safeguards. 
\begin{center}\textit{"There could be data concerns depending on the document being uploaded. Personal documents would matter more than general content."}[P2]\end{center}
Without strict guidelines, there is a risk of sensitive student data being compromised. As AI adoption grows in education, policies that address ethical data use and privacy protections will be necessary to build trust among users.

Potential misuse of AI tools in education was another key concern. While AI can enhance learning, it also presents opportunities for academic dishonesty. 
\begin{center}\textit{"There might be misuse, like copying answers, so monitoring is important."}[P12]\end{center}
\begin{center}\textit{"Using tools like ChatGPT for project work directly might hinder students' ability to broaden their horizons by researching through multiple sources."}[P3]\end{center}
Establishing ethical AI usage policies in schools, alongside digital literacy programs, can help mitigate these risks and ensure students engage with AI responsibly.

Cultural resistance to AI adoption also surfaced as a barrier. Some participants noted that government school teachers, in particular, were unfamiliar with GenAI, leading to skepticism about its role in education. 
\begin{center}\textit{"Teachers often resist AI tools like ChatGPT, seeing them as a shortcut or a form of cheating rather than an educational aid."}[P22]\end{center}
\begin{center}\textit{"Resources alone aren't enough—there must be awareness programs to help teachers and students embrace new technologies."}[P15]\end{center}
The Diffusion of Innovations Theory \cite{rogers2003diffusion} explains that technology adoption follows a pattern where initial resistance is common, but with proper awareness and training, more users gradually accept it. Structured teacher training and awareness programs could help alleviate concerns and encourage AI integration into classrooms.

These concerns underscore the need for a balanced approach in implementing AI in rural education. While AI can be a powerful educational tool, safeguards against overdependence, privacy risks, and misuse must be in place. Additionally, targeted awareness efforts can help shift cultural perceptions, ensuring that AI adoption is gradual, ethical, and aligned with the best interests of students and educators.

\subsection{Community Reactions to AI}
\label{sec:community}

Community perceptions of AI in rural education vary significantly, with reactions ranging from optimism to skepticism. Participants highlighted both the enthusiasm for AI's potential in improving educational outcomes and concerns about accessibility, awareness, and adaptation among teachers and parents.

When asked about AI's reception in rural communities, participants reported mixed reactions. Some recognized its potential to enhance learning and expand opportunities, while others were skeptical, seeing it as a distraction or an unfamiliar technology. 
\begin{center}\textit{"There will be mixed reactions from communities. Some may see it as a distraction, while others might recognize its value for learning."}[P6]\end{center}
Others highlighted positive initiatives, such as AI-driven educational programs by volunteers:
\begin{center}\textit{"We at NSS have collaborated with PARC to teach selected students web development and Excel basics, leading to a certification."}[P19]\end{center}
These responses indicate that while AI has begun making inroads in rural education, broader acceptance depends on continued exposure and demonstrable benefits.

Awareness and training were frequently cited as essential for AI adoption. Many participants emphasized the need for structured workshops and guidance, particularly for students in higher grades. 
\begin{center}\textit{"We invited students from classes 9 to 12 for a two-day workshop at IIIT, providing them with career counseling sessions and discussing their aspirations."}[P2]\end{center}
\begin{center}\textit{"Workshops would be very helpful in ensuring AI tools are used effectively and responsibly."}  
 [P17]\end{center}
Without adequate training, students and teachers may struggle to utilize AI tools to their full potential. The Technology Readiness Index \cite{parasuraman2000technology} suggests that successful technology adoption depends on users' readiness and confidence, underscoring the importance of well-structured AI awareness initiatives in rural schools. Participants suggested that NGOs and college volunteers play a crucial role in bridging this gap by introducing AI tools in schools and helping both teachers and students gain confidence in their usage.

Parental support emerged as a key factor influencing AI adoption in rural education. Some parents see AI as a valuable resource for their children's future, while others are more hesitant. 
\begin{center}\textit{"Parents and the community would see AI as a way to ensure a brighter future for their children."} [P14]\end{center}
However, many students remain unaware of career paths or how AI can support their aspirations, reinforcing the need for parental engagement. 
\begin{center}\textit{"We must hold demo classes and involve parents to show how helpful AI can be in educating their kids."} [P21]\end{center}
If parents gain firsthand experience with AI's benefits, they may be more inclined to support its integration into their children's education. Effective AI integration requires not only educating students but also building trust among families who play a crucial role in shaping educational priorities.

Teachers play a critical role in shaping AI's success in rural classrooms, yet their adaptation to AI remains uneven. Participants proposed that some educators will welcome AI as a support tool, while others might resist its implementation, fearing it might diminish traditional teaching methods. 
\begin{center}\textit{"Successful integration would start with educating teachers and students about Gen AI's potential. They need to understand that it's there to support them, not replace them."} [P6]\end{center}
The Diffusion of Innovations Theory \cite{rogers2003diffusion} suggests that technology adoption follows a pattern where early adopters embrace innovation while others remain skeptical until they observe clear benefits. To ensure AI adoption is widely accepted, structured teacher training and awareness campaigns must address these concerns and demonstrate AI's role as a complementary tool rather than a replacement.

\subsection{The Future of AI in Education}
\label{sec:policy}

As AI continues to evolve, its role in education is expected to expand significantly. Participants expressed both excitement and caution about the long-term integration of AI in rural schools, highlighting its potential for transforming learning experiences while emphasizing the need for careful implementation.

When asked about the future of AI in rural education, participants were optimistic about its potential to enhance both teaching and learning experiences. Among the 23 participants, 8 mentioned that AI could be a valuable educational aid if integrated properly. 
\begin{center}\textit{"If tools like ChatGPT were made available and if teachers and students were trained on using them, it could make a significant difference."} [P18]\end{center}
\begin{center}\textit{"Generative AI could assist teachers with lesson planning, student feedback, and administrative tasks."} 
 [P19]\end{center}
These insights suggest that while AI will not replace traditional education models, it could serve as an essential tool to enhance efficiency and accessibility. The Technology Readiness Index \cite{parasuraman2000technology} suggests that successful AI adoption depends on both access to technology and user preparedness, highlighting the importance of training initiatives in rural schools.

Participants also shared long-term visions for AI's role in shaping the future of education. Some saw AI as a way to close the educational divide between urban and rural areas. 
\begin{center}\textit{"With AI, they could learn better, which would improve the overall quality of education and help narrow the gap between rural and urban areas."} [P1]\end{center}
\begin{center}\textit{"In the long run, AI in education would be a great investment because it encourages both self-learning and creativity."} [P13]\end{center}
Future AI integration should align with Vygotsky's Zone of Proximal Development \cite{vygotsky1978mind}, ensuring that AI acts as a scaffold to help students reach their full potential rather than replacing human instruction.

The impact of AI on traditional teaching methods remains a subject of debate. While AI offers advantages such as personalized learning and automated administrative support, participants stressed the continued importance of human interaction in education. 
\begin{center}\textit{"While AI will be important, I think the human touch is still very much required in education."} [P15]\end{center}
\begin{center}\textit{"With the right framework, students could learn independently and teachers could focus on conceptual clarity."} [P23]\end{center}
The integration of AI will likely reshape teaching practices, but educators will continue to play an essential role in providing mentorship, emotional support, and critical thinking skills.

Blended learning models emerged as a preferred approach for integrating AI into rural education. Participants emphasized that AI should complement rather than replace existing teaching methods. 
\begin{center}\textit{"AI-driven education must be a mix of structured guidance and free exploration for the best results."} [P17]\end{center}
\begin{center}\textit{"If AI tools were designed in a way that makes learning more interactive and appealing, it might encourage students."} [P2]\end{center}
The success of AI in education will depend on balancing technological advancements with personalized, teacher-led instruction to create an effective learning environment.

\subsection{Visualization and Creativity in Learning}
\label{sec:visual}

AI has the potential to revolutionize education by enhancing visualization and fostering creativity among students. Participants highlighted the importance of using AI-generated visuals, interactive content, and real-life applications to improve student engagement and comprehension.

When asked about AI's role in enhancing visualization, participants emphasized its ability to generate interactive content that makes abstract concepts more accessible. Among the 23 participants, 12 mentioned that AI-generated videos, images, and simulations could improve students' understanding of complex subjects. 
\begin{center}\textit{"I imagine a setup where a tool generates relevant images, graphs, or videos as the teacher speaks, making lessons more interactive and visually engaging."} [P22]\end{center}
\begin{center}\textit{"AI-generated videos help students understand better because visuals stick in their memory."} [P17]\end{center}
Cognitive Load Theory \cite{sweller1988cognitive} suggests that students learn best when complex information is presented through multiple modalities, reinforcing AI's potential to enhance STEM education by reducing cognitive overload and improving retention.

AI also holds promise for fostering creativity among students by enabling personalized and exploratory learning. Some participants expressed excitement about AI's potential to encourage creative problem-solving, while others warned of potential drawbacks. 
\begin{center}\textit{"Generative AI could enhance creativity in students by allowing them to explore and understand concepts practically."} [P3]\end{center}
\begin{center}\textit{"At the same time, I feel AI sometimes kills my creativity, as I depend on it to find answers instead of thinking on my own."} [P9]\end{center}
The Self-Determination Theory \cite{ryan2000self} highlights the importance of autonomy and engagement in learning, suggesting that AI should be implemented in ways that encourage students to think independently rather than becoming overly reliant on automated solutions.

Participants highlighted the importance of connecting classroom learning with real-world applications to improve student motivation. 
\begin{center}\textit{"AI can create videos and content, especially for science, as it helps connect concepts to real life, making them easier to understand."} [P18]\end{center}
\begin{center}\textit{"Science concepts, like acids and bases, can be connected to everyday life through videos, making them easier to grasp."} [P8]\end{center}
AI-driven tools can bridge the gap between theoretical knowledge and practical application, making learning more meaningful for students in rural areas.

Interactive content emerged as a key factor in AI's potential to improve student engagement. Participants noted that AI could be particularly effective in making subjects like science and mathematics more accessible through gamification and hands-on learning. 
\begin{center}\textit{"When I was teaching, I discovered an application that gave students questions tailored to their skill level, which made learning interactive and enjoyable for young kids."} [P7]\end{center}
\begin{center}\textit{"If students were made aware of tools like ChatGPT, they could use them easily and benefit significantly."} [P20]\end{center}
By creating adaptive and engaging learning environments, AI can help students develop problem-solving skills and curiosity.

\vspace{1em}

The findings of this study highlight both the potential and the challenges of integrating AI in rural K-12 education. While AI offers promising solutions to bridge learning gaps, enhance visualization, and personalize education, its success is contingent on addressing infrastructure limitations, teacher training gaps, and community skepticism. Participants acknowledged AI's role as a valuable supplementary tool but emphasized that it cannot replace traditional teaching methods. Concerns about overreliance on AI, data privacy, and potential misuse underscore the need for ethical implementation strategies. The study also revealed that teacher and student awareness programs, parental engagement, and blended learning approaches are crucial to ensuring AI's effective adoption. By fostering digital literacy and providing adequate training, rural education can gradually leverage AI to improve learning outcomes. However, sustainable implementation requires policy support, financial investment, and culturally sensitive AI tools.

\section{Discussion}
\label{sec:discussion}

The integration of AI in rural K-12 education is positioned at the intersection of technological innovation and systemic educational challenges. While AI tools have demonstrated potential to personalize learning, assist teachers, and bridge urban-rural educational disparities, their successful adoption in rural schools depends on several critical factors, including infrastructure readiness, teacher training, community acceptance, and ethical considerations. Previous technology-for-education initiatives, such as efforts to provide free Internet connectivity to underserved populations \cite{internetorg2020}, have often struggled to achieve long-term impact due to underlying socio-economic and infrastructural barriers. AI adoption in rural education faces similar risks if these foundational challenges are not addressed.

Our findings indicate that while student volunteers generally perceive AI as a useful tool for enhancing education, skepticism persists regarding over-reliance, digital literacy gaps, and accessibility constraints. Additionally, systemic barriers such as poor infrastructure, limited professional development opportunities, and parental resistance hinder AI's effective deployment. Beyond practical considerations, concerns about data privacy, cultural acceptance, and long-term sustainability of AI tools also emerged as important factors. The following discussion addresses our research questions by examining four key themes: (1) the perceptions and attitudes of student volunteers toward AI in rural education, (2) the structural and social barriers limiting AI adoption in rural schools, (3) the potential of AI to create personalized learning experiences that bridge educational gaps, and (4) the ethical considerations and policy implications for responsible AI implementation. 

\subsection{Student Volunteers’ Perceptions of AI in K-12 Education}
\label{sec:disc_perceptions}

Our findings indicate that student volunteers in rural schools exhibit both optimism and skepticism regarding AI’s role in the classroom. While many recognize AI’s ability to support personalized and context-sensitive learning, automate repetitive tasks, and enhance student engagement, concerns persist about over-reliance, student readiness, and alignment with traditional teaching methods.

\textbf{How do Student Volunteers perceive AI’s potential in education?} They largely view AI as a supportive rather than a transformative tool, capable of enhancing classroom instruction rather than replacing teachers. AI-driven content, student-specific assessments, and automation of administrative tasks are seen as valuable additions to the education system. However, past attempts to introduce technology in education, such as efforts to distribute low-cost laptops for digital learning, have shown that technology alone does not automatically translate to better learning outcomes \cite{toyama2011}. AI, like previous innovations, risks being ineffective if it is not properly integrated into pedagogical strategies that consider the realities of rural education.

Skepticism arises from concerns that AI might discourage independent thinking among students, who may become passive learners if AI tools are not used in a structured manner. Similar critiques were raised during the early adoption of educational television programs and online learning initiatives, where studies found that without active teacher intervention, students engaged in passive content consumption rather than interactive learning and knowledge building \cite{selwyn2016}.

\textbf{What factors influence teacher's acceptance or skepticism toward AI?} A major factor shaping AI adoption is the lack of familiarity and training among rural educators. While some teachers have experimented with digital tools, many remain unaware of AI’s capabilities or lack access to structured training programs. This reflects a broader challenge seen in past digital education initiatives, where teachers were expected to integrate technology without sufficient technical support or professional development \cite{zhao2003}. Without AI literacy programs tailored to rural schools, educators may find it difficult to incorporate AI tools effectively.

Additionally, high workloads and limited school resources make it challenging for teachers to explore new technologies. AI implementation adds to their responsibilities, requiring time for training and classroom adaptation. Similar constraints were observed in previous educational technology interventions, where teacher burnout and resource limitations led to low adoption rates despite initial enthusiasm \cite{cuban2001}.

Another critical concern is AI’s relevance in low-resource settings, where many students struggle with literacy and numeracy basics. If AI-driven learning tools assume a baseline level of digital literacy, they may fail to meet the needs of students who lack foundational academic skills. To ensure accessibility, AI must first gauge students’ individual learning levels and then provide targeted support, allowing each student to progress at their own pace. Previous studies on the digital divide highlight that introducing advanced technology without addressing fundamental learning barriers can widen educational inequalities rather than reducing them \cite{warschauer2004}.

\subsection{Barriers to AI Adoption in Rural Schools}
\label{sec:disc_barriers}

Our findings reveal that while AI has the potential to enhance education in rural schools, several structural, pedagogical, and socio-cultural barriers hinder its adoption. The most significant challenges include inadequate infrastructure, lack of teacher training, financial constraints, and parental skepticism. These barriers not only limit access to AI-driven tools but also affect the willingness and capacity of educators to integrate AI into their teaching practices.

\textbf{How does infrastructure affect AI adoption?} A major obstacle to AI implementation in rural schools is the lack of stable internet connectivity, electricity, and digital devices. While urban schools are progressively incorporating AI-based learning platforms, rural institutions struggle with basic technological infrastructure, making AI-enhanced education largely inaccessible. Similar challenges have been observed in previous efforts to expand digital learning in underserved regions, where studies found that poor infrastructure often rendered technological interventions ineffective, despite their potential benefits. To address this, mobile-based AI learning applications that require minimal connectivity and enable offline learning could provide an alternative solution.

Even in cases where digital tools are introduced, unreliable power supply and limited internet access significantly disrupt AI-based learning. Prior research on ICT adoption in education suggests that without a stable technological foundation, digital learning initiatives risk being unsustainable, leading to inconsistent usage and eventual abandonment \cite{donner2015}. These findings indicate that AI cannot be meaningfully integrated into rural schools without first addressing fundamental infrastructure gaps.

\textbf{How does teacher training impact AI adoption?} The effectiveness of AI in education largely depends on the ability of teachers to use and integrate AI-driven tools effectively. However, many rural educators lack formal training in AI-based teaching methods and are unfamiliar with AI’s potential applications in education. Studies on digital literacy among teachers indicate that successful technology adoption requires more than access to tools—it depends on educators’ confidence and competence in using them \cite{ertmer2010}. Providing structured AI training, with a focus on adaptable and individualized instruction, could significantly enhance AI’s impact in classrooms.

Previous research on large-scale digital education initiatives has shown that introducing technology without adequate teacher preparation leads to resistance or ineffective implementation \cite{selwyn2011}. In many cases, rural educators express frustration when expected to integrate new technology without sufficient professional development, further widening the gap between policy expectations and classroom realities. Without dedicated AI training and ongoing support, teachers are unlikely to feel equipped to integrate AI effectively into their instructional practices.

\textbf{What role does parental skepticism play in AI adoption?} Parental attitudes toward AI-driven education significantly impact its adoption in rural schools. Many families remain skeptical about AI’s role in learning, viewing it as either unnecessary or potentially harmful. Concerns about excessive screen time, AI’s influence on traditional learning methods, and the potential reduction of human interaction shape parental resistance to AI integration. Similar skepticism was observed in past efforts to introduce digital learning tools in low-income communities, where parents often perceived technology as a distraction rather than an educational aid \cite{selwyn2016}.

Additionally, limited exposure to AI-driven education contributes to mistrust and uncertainty among parents. Studies on technology acceptance in education have highlighted that without parental engagement and awareness, new learning technologies face significant resistance, particularly in communities where traditional teaching methods are deeply valued \cite{venkatesh2003}. To foster AI adoption in rural schools, educators and policymakers must actively involve parents in discussions about AI’s benefits, ensuring that AI is seen as a tool to enhance—not replace—traditional education.

\textbf{How do financial constraints impact AI accessibility?} Financial limitations further restrict AI adoption in rural schools, as many institutions lack the budget to invest in AI-driven educational technology. Even when AI-based learning tools are made available, families may not have the financial means to provide students with personal digital devices or reliable internet access for learning outside the classroom. Research on the digital divide has shown that cost remains one of the most significant barriers to technology adoption in education, particularly in underprivileged communities \cite{warschauer2004}.

Similar challenges have been encountered in initiatives aimed at distributing low-cost digital devices to students, where hardware availability did not always translate into improved learning outcomes due to the lack of ongoing financial and technical support \cite{toyama2011}. Without long-term investments in both infrastructure and affordability, AI-driven education risks being concentrated in well-funded urban schools, further exacerbating existing educational inequalities.

\subsection{AI’s Role in Personalized Learning and Bridging the Education Gap}
\label{sec:disc_personalized}

Our findings suggest that AI has the potential to improve personalized learning, adaptive instruction, and equitable access to quality education for rural students. However, its effectiveness depends on how well AI tools are designed to address the specific needs of rural learners. While AI-driven education could help bridge the rural-urban education gap, challenges such as low digital literacy, language barriers, and unequal access to AI-powered resources must be addressed to ensure its benefits are fully realized.

\textbf{How can AI personalize learning in rural schools?} One of AI’s most significant advantages in education is its ability to adapt to individual student needs, providing tailored instruction that adjusts to different learning levels. In classrooms where students exhibit diverse learning abilities, AI-powered platforms can offer real-time feedback, personalized assessments, and adaptive learning pathways. Research on digital learning tools has shown that adaptive AI-based instruction can help struggling students reinforce foundational skills while allowing advanced learners to progress at a faster pace \cite{rose2010}.

However, the challenge in rural settings lies in ensuring that AI-based learning tools account for foundational skill gaps. If AI models assume a baseline level of digital literacy, students who lack essential reading, writing, and numeracy skills may struggle to engage with AI-generated content effectively. Studies on technology-enhanced education indicate that personalized learning is most effective when supported by human educators who can scaffold student interactions with AI tools \cite{luckin2018}. Without structured teacher involvement, AI-driven learning risks being less accessible to students with weaker academic foundations.

Another critical factor is language accessibility. Many AI-powered learning tools are designed primarily for English or widely spoken national languages, which may not align with the regional languages spoken in rural communities. Research on AI in multilingual education suggests that without localized content, AI-based learning tools may fail to reach a significant portion of students in non-English-speaking regions \cite{strigel2012}. To maximize the impact of AI-driven personalized learning, AI models must be adapted to include regional languages and culturally relevant educational materials.

\textbf{Can AI help bridge the rural-urban education divide?} AI has the potential to expand access to high-quality learning resources for students in rural areas, helping to reduce disparities in education. In low-resource environments where teacher shortages, outdated curricula, and limited access to learning materials hinder student progress, AI-driven tools can supplement traditional instruction by providing interactive content, automated tutoring, and access to global knowledge networks. Studies on digital education in underserved regions suggest that AI-powered platforms can offer a cost-effective means of delivering standardized, high-quality instruction to students who lack access to trained educators.

However, while AI has the potential to narrow the learning gap between urban and rural students, its success depends on infrastructure availability, digital literacy, and equitable access to technology. Previous research on e-learning disparities highlights that students in well-connected urban areas are more likely to benefit from AI-enhanced education than their rural counterparts, exacerbating existing educational inequalities \cite{warschauer2004}.

To ensure that AI-driven education serves as a bridging mechanism rather than a dividing factor, policymakers and educational institutions must invest in infrastructure, provide AI literacy training for teachers, and develop AI-powered learning tools that cater to the specific needs of rural students. Efforts to integrate AI into rural education should focus on blended learning models, where AI supplements teacher-led instruction rather than replacing it. Prior studies have emphasized that AI is most effective when combined with human interaction, allowing teachers to use AI insights to tailor instruction and provide targeted support \cite{luckin2018}.

Another challenge in bridging the education gap is ensuring students in rural schools have access to AI-driven learning outside the classroom. While mobile phones are becoming more prevalent in rural communities, their use for educational purposes remains limited, with students primarily using them for entertainment or informal learning. Similar concerns were raised in studies examining low-cost tablet distribution programs, where researchers found that without structured guidance and learning incentives, students did not engage with educational content meaningfully \cite{toyama2011}. Expanding AI-based learning opportunities beyond school settings requires community engagement, affordable access to digital devices, and initiatives that promote AI-driven learning as a viable supplement to traditional education.

\subsection{Ethical and Practical Concerns in AI Adoption}
\label{sec:disc_ethics}

Our findings indicate that while AI presents significant opportunities for improving education in rural schools, its integration raises ethical, pedagogical, and practical concerns that must be addressed. Key issues include data privacy, algorithmic bias, over-reliance on AI-driven learning, and sustainability challenges. These concerns highlight the need for responsible AI implementation strategies that prioritize transparency, equity, and long-term viability in rural education settings.

\textbf{What are the key ethical concerns surrounding AI in rural education?} One of the primary ethical concerns is student data privacy. AI-driven educational platforms rely on data collection and analysis to personalize learning experiences, raising questions about who has access to this data, how it is stored, and whether students and teachers are adequately informed about data usage policies. Past research on digital privacy in education suggests that students in low-resource settings are particularly vulnerable to data exploitation, as they may lack awareness of their digital rights and protections. Without strong data governance frameworks and clear regulatory guidelines, AI-based learning tools risk exposing students to privacy breaches and potential misuse of personal information.

Another ethical challenge is algorithmic bias, where AI-driven educational tools may reinforce existing inequalities rather than mitigating them. Studies on AI fairness indicate that AI models trained on urban-centric or Western educational datasets may fail to accurately reflect the learning needs, linguistic diversity, and socio-cultural context of rural students \cite{baker2021}. This can lead to content mismatches, biased assessment results, and a lack of inclusivity in AI-generated educational resources. Addressing this issue requires deliberate efforts to develop AI models that are representative of diverse learning environments, incorporating localized content and culturally relevant materials.

\textbf{How can AI be implemented responsibly to avoid over-reliance?} Over-reliance on AI-driven education presents another significant concern. While AI-powered learning tools can enhance instruction, excessive dependence on AI may reduce students’ ability to engage in independent problem-solving and critical thinking. Similar concerns were raised in studies on educational automation, where researchers found that students who frequently relied on AI-generated answers exhibited lower retention rates and reduced analytical skills compared to those who actively engaged in human-led discussions \cite{brougham2018}.

Additionally, teachers may face pressure to rely on AI for instructional decision-making, which could lead to reduced autonomy in pedagogical choices. Research on AI adoption in classrooms suggests that when educators perceive AI as a rigid system rather than a flexible support tool, they are less likely to integrate it in meaningful ways \cite{selwyn2016}. To mitigate these risks, AI adoption should be guided by structured frameworks that ensure AI serves as an enhancement to—rather than a replacement for—traditional learning methods.

\textbf{What are the sustainability challenges of AI integration in rural education?} The long-term sustainability of AI in rural education depends on continued investment in digital infrastructure, teacher training, and policy support. Many past technology-driven education initiatives failed due to a lack of ongoing technical support, financial backing, and adaptability to local conditions \cite{toyama2011}. Without proper maintenance and iterative improvements, AI tools introduced in rural schools may become obsolete, underutilized, or completely abandoned.

Another sustainability challenge is the cost of AI-driven education. While some AI-powered tools are freely available, many advanced learning platforms require paid subscriptions, device compatibility, and stable internet connectivity, making them inaccessible to students in lower-income rural communities. Prior studies on digital divide issues emphasize that when educational technologies are not financially sustainable, their adoption remains limited to well-funded institutions, further widening educational disparities \cite{warschauer2004}. Ensuring affordable AI solutions, open-access educational tools, and long-term government or NGO-backed support will be critical in sustaining AI-driven learning in rural areas.

AI’s effectiveness in rural education also hinges on how well it aligns with local educational policies and assessment frameworks. If AI-based learning methods do not integrate with existing curricula, standardized testing requirements, and teacher evaluation systems, their adoption may remain superficial rather than transformative. Past research on ed-tech policy implementation highlights that a lack of alignment between technology initiatives and formal education policies often leads to fragmented and inconsistent adoption patterns. Successful AI integration in rural education must therefore involve collaboration between policymakers, educators, and technology developers to ensure AI tools complement national education goals.

\section{Conclusion}

This study explored the perceptions of volunteer educators on the integration of Large Language Models (LLMs) in rural K-12 education in India. While the participants expressed cautious optimism about the potential of AI to personalize learning, reduce teacher workload, and enhance student engagement, they also articulated a number of valid concerns. Supportive voices in the study emphasized how AI could fill critical learning gaps, especially by offering scalable and adaptive instruction in regions facing teacher shortages and foundational learning deficits. They highlighted that with appropriate teacher training and contextual localization, AI-powered tools could democratize access to quality education and inspire rural students to engage more deeply with their academic and career aspirations.

However, not all participants were unequivocally enthusiastic. Several volunteers voiced concerns about the risk of over-reliance on AI, loss of student creativity, language barriers, and resistance from teachers and parents. Ethical anxieties around data privacy, as well as fears that AI might deskill teachers or widen the rural-urban digital divide, emerged consistently. These findings underscore that AI must be introduced not as a disruptive replacement, but as a carefully scaffolded supplement to human-led teaching practices in rural classrooms.

In response to these insights, our team is currently developing an \textbf{AI-driven mobile application} tailored specifically for the needs of students in rural India. The app will cover \textbf{core NCERT curriculum content from Grades 3 to 12}, and will feature both \textbf{Theory Revision Mode} and \textbf{Practice Mode}. Students will be able to revise key concepts, explore simplified summaries, and receive customized suggestions based on their learning progress. Additionally, the app will help users \textbf{find, understand, and solve NCERT textbook problems}, offering step-by-step explanations in regional languages wherever possible.

Designed with low-bandwidth compatibility, image and voice-based search, the app aims to accommodate students with limited digital literacy and infrastructural access. Our vision is to build not just a repository of content, but a \textit{conversational learning assistant} that empowers rural students to learn at their own pace, with minimal barriers. Grounded in the findings of this research, the app aspires to bring the promise of LLMs directly into the hands of learners who need them most.

Future work will involve field-testing the app in rural classrooms and incorporating feedback from teachers, students, and parents to ensure that the technology remains pedagogically sound, culturally relevant, and ethically responsible. As AI continues to evolve, its role in education must be guided not only by innovation but also by inclusivity and empathy.

\bibliography{sample}

\section*{Author contributions statement}
H.G. led the analysis and original draft writing. 
All authors conceptualised, collected data, conducted analysis, and reviewed the manuscript. 

\section*{Competing Interests}
The authors declare no competing interests. 

\section*{Additional information}
Correspondence and requests for material should be addressed to V.R. (veena.r@pilani.bits-pilani.ac.in). The ethics committee that approved this study was the Institutional Human Ethics Committee (IHEC), BITS Pilani, Rajasthan, India.

\section*{Data Availability Statement}
The datasets used and/or analyzed during the current study are available from the corresponding author upon request.

\newpage
\section*{Appendix A: INTERVIEW PROTOCOL}

\begin{itemize}
    \item \textbf{General Background on Education and Technology}
    \begin{itemize}
        \item Can you describe the current state of K-12 education in rural areas you are familiar with?
        \item What are the main challenges students and teachers face in rural schools?
        \item In your experience, how has technology been used (or not) in rural education so far?
    \end{itemize}

    \item \textbf{Perceptions of GenAI}
    \begin{itemize}
        \item Have you heard of or used tools like chatbots or AI-based language models before? If so, what was your experience?
        \item How do you think GenAI could be useful, if at all, in improving education quality in rural areas?
        \item What concerns or challenges do you foresee in using AI tools like GenAI in rural schools?
    \end{itemize}

    \item \textbf{Practical Applications and Potential Impact}
    \begin{itemize}
        \item What kinds of educational content or support do you think GenAI could provide to students in rural areas?
        \item Do you think GenAI could help reduce the gap between rural and urban education quality? If so, how?
        \item In what ways could GenAI assist teachers in rural schools?
    \end{itemize}

    \item \textbf{Barriers to Implementation}
    \begin{itemize}
        \item What do you think could be the major barriers to adopting AI tools like GenAI in rural schools?
        \item What infrastructure (e.g., internet, devices) do you think is needed to make GenAI usable in rural schools?
        \item How do you think we can address concerns around the cost and accessibility of these technologies?
    \end{itemize}

    \item \textbf{Volunteer and Community Perspectives}
    \begin{itemize}
        \item How do you think the community, parents, and local teachers might react to the use of GenAI in education?
        \item What role could volunteers or NGOs like NSS play in introducing and supporting GenAI in rural education?
    \end{itemize}

    \item \textbf{Ethical and Cultural Considerations}
    \begin{itemize}
        \item Are there any ethical concerns you think should be considered when using AI tools in education, particularly in rural settings?
        \item How important is it for these tools to be available in local languages or dialects?
    \end{itemize}

    \item \textbf{Future Outlook}
    \begin{itemize}
        \item Looking forward, how do you see the role of technology, and specifically GenAI, evolving in rural education?
        \item What would a successful integration of GenAI in rural schools look like to you?
    \end{itemize}
\end{itemize}

\newpage
\section*{Appendix B: Codebook}
\begin{table}[h!]
\centering
\caption{CODEBOOK AND THEMES FROM QUALITATIVE ANALYSIS}
\begin{tabular}{|l|l|r|}
\hline
\textbf{Theme} & \textbf{Code} & \textbf{No. of Quotes} \\
\hline
Challenges in Rural Education & Infrastructure Deficits & 16 \\
 & Teacher Quality & 15 \\
 & Student Engagement Issues & 22 \\
 & Parental Attitudes & 9 \\
 & \textbf{Total} & \textbf{62} \\
\hline
The Role of Technology in Education & Current Use of Technology & 21 \\
 & Potential of Technology & 28 \\
 & Blending Technology with Traditional Methods & 17 \\
 & Visualization Benefits & 11 \\
 & \textbf{Total} & \textbf{77} \\
\hline
Personalized Learning with GenAI & Catering to Individual Needs & 13 \\
 & Support for Teachers & 18 \\
 & Bridging Knowledge Gaps & 12 \\
 & Reducing Urban-Rural Gaps & 9 \\
 & \textbf{Total} & \textbf{52} \\
\hline
Barriers to GenAI Adoption & Cost and Accessibility & 11 \\
 & Language Barriers & 8 \\
 & Infrastructure Challenges & 8 \\
 & Teacher Training & 9 \\
 & \textbf{Total} & \textbf{36} \\
\hline
Ethical and Practical Concerns & Overdependence on AI & 8 \\
 & Data Privacy Issues & 1 \\
 & Potential Misuse & 4 \\
 & Cultural Resistance & 5 \\
 & \textbf{Total} & \textbf{18} \\
\hline
Community Reactions to AI & Mixed Reactions & 4 \\
 & Awareness and Training & 7 \\
 & Parental Support & 3 \\
 & Teacher Adaptation & 4 \\
 & \textbf{Total} & \textbf{18} \\
\hline
The Future of AI in Education & Dream of Integration & 8 \\
 & Long-Term Vision & 7 \\
 & Impact on Traditional Teaching & 6 \\
 & Blended Learning Models & 8 \\
 & \textbf{Total} & \textbf{29} \\
\hline
Visualization and Creativity in Learning & Enhancing Visualization & 12 \\
 & Fostering Creativity & 9 \\
 & Real-Life Connections & 8 \\
 & Interactive Content & 7 \\
 & \textbf{Total} & \textbf{36} \\
\hline
\end{tabular}
\end{table}

\end{document}